\documentclass[%
reprint,
superscriptaddress,
preprintnumbers,
nofootinbib,
amsmath,amssymb,
aps,
 floatfix
]{revtex4-2}

\usepackage{graphicx}
\usepackage{dcolumn}
\usepackage{bm}
\usepackage{siunitx}
\usepackage[version=4]{mhchem}
\usepackage[caption=false]{subfig}
\newcommand{\phantomsubfloat}[1]{
    {
        \captionsetup[subfigure]{labelformat=empty}
        \subfloat[][]{#1}
        \vspace{-5pt}
    }%
}

\usepackage{amsmath}
\usepackage{collcell}
\usepackage{makecell}
\usepackage{comment}
\usepackage[normalem]{ulem}
\usepackage{hyperref}
\usepackage{cleveref}
\crefname{section}{sec.}{secs.}
\usepackage{orcidlink}

\usepackage[mathlines,pagewise]{lineno}

\usepackage{titlesec}
\titleformat{\section}{\centering\normalfont\bfseries}{\thesection}{.5em}{\vspace{.5ex}}

\begin{document}

\title{R2D2 -- An equivalent-circuit model that quantitatively describes domain wall conductivity in ferroelectric \texorpdfstring{\ce{LiNbO3}}{LiNbO3}}  
                                                       
\author{Manuel~Zahn\,\orcidlink{0000-0003-0739-3049}}
\affiliation{Institute of Applied Physics, Technische Universit\"at Dresden, 01062 Dresden, Germany}
\affiliation{Experimental Physics V, Center for Electronic Correlations and Magnetism, University of Augsburg, 86159~Augsburg, Germany}

\author{Elke~Beyreuther\,\orcidlink{0000-0003-1899-603X}}%
 \email{elke.beyreuther@tu-dresden.de}
\affiliation{Institute of Applied Physics, Technische Universit\"at Dresden, 01062 Dresden, Germany}

\author{Iuliia~Kiseleva\, \orcidlink{0009-0002-5435-056X}}
\affiliation{Institute of Applied Physics, Technische Universit\"at Dresden, 01062 Dresden, Germany}

\author{Ahmed~Samir~Lotfy\,\orcidlink{0000-0001-9874-8460}}%
\affiliation{Department of Materials, ETH Z\"urich, 8093 Zürich, Switzerland}

\author{Conor~J.~McCluskey\,\orcidlink{0000-0001-8857-5105}}
\affiliation{Center for Quantum Materials and Technologies, School of Mathematics and Physics, Queen's University Belfast, Northern Ireland}

\author{Jesi~R.~Maguire\,\orcidlink{0000-0003-2362-8717}}
\affiliation{Center for Quantum Materials and Technologies, School of Mathematics and Physics, Queen's University Belfast, Northern Ireland}

\author{Ahmet~Suna\,\orcidlink{0000-0001-8141-8617}}
\affiliation{Center for Quantum Materials and Technologies, School of Mathematics and Physics, Queen's University Belfast, Northern Ireland}

\author{Michael~R\"using\,\orcidlink{0000-0003-4682-4577}}
\affiliation{Institute of Applied Physics, Technische Universit\"at Dresden, 01062 Dresden, Germany}
\affiliation{Integrated Quantum Optics, Institute for Photonic Quantum Systems (PhoQS), Paderborn University, 33098 Paderborn, Germany}

\author{J.~Marty~Gregg\,\orcidlink{0000-0002-6451-7768}}
\affiliation{Center for Quantum Materials and Technologies, School of Mathematics and Physics, Queen's University Belfast, Northern Ireland}

\author{Lukas~M.~Eng\,\orcidlink{0000-0002-2484-4158}}
\affiliation{Institute of Applied Physics, Technische Universit\"at Dresden, 01062 Dresden, Germany}
\affiliation{ct.qmat: Dresden-W\"urzburg Cluster of Excellence--EXC 2147, Technische Universit\"at Dresden, 01062 Dresden, Germany}

\date{\today}

\begin{abstract}
    Ferroelectric domain wall (DW) conductivity (DWC) can be attributed to two separate mechanisms: (a) the injection/ejection of charge carriers across the Schottky barrier formed at the (metal\nobreakdash-){\allowbreak}electrode-DW junction and (b) the transport of those charge carriers along the DW. Current-voltage (IU) characteristics, recorded at variable temperatures from \ce{LiNbO3} (LNO) DWs, are clearly able to differentiate between these two contributions. Practically, they allow us here to directly quantify the physical parameters relevant for the two mechanisms (a) and (b) mentioned above. These are, e.g., the resistance of the DW, the saturation current, the ideality factor, and the Schottky barrier height of the electrode/DW junction. Furthermore, the activation energies needed to initiate the thermally-activated electronic transport along the DWs, can be extracted. In addition, we show that electronic transport along \ce{LiNbO3} DWs can be elegantly viewed and interpreted in an adapted semiconductor picture based on a double-diode/double-resistor equivalent circuit model, the R2D2 model. Finally, our R2D2 model was checked for its universality by successfully fitting the IU curves of not only z-cut LNO bulk DWs, but equally of z-cut thin-film LNO DWs, and of x-cut thin-film DWs as reported in literature. 
\end{abstract}


\maketitle

\section{Introduction}
\label{sec:intro}

Since the early prediction of enhanced electrical conductivity along charged ferroelectric domain walls in the 1970s \cite{vul_encountering_1973}, the intriguing phenomenon of domain wall conductivity (DWC) has been reported in a number of ferroelectric materials during the last decade, which opens a unique perspective for designing integrated functional nanoelectronic elements \cite{catalan_domain_2012,meier_functional_2015}. The enormous scientific interest is reflected in several review articles treating fundamental \cite{sluka_charged_2016,bednyakov_physics_2018,nataf_domain-wall_2020} and technological \cite{sharma_functional_2019,meier_ferroelectric_2021,sharma_roadmap_2022} challenges in understanding and exploiting this type of quasi-2-dimensionally confined electronic transport, which competes with other highly topical low-dimensional electronic systems like graphene, oxide interfaces, or heterointerfaces of classical semiconductors. Notably, DW based approaches bear the unique possibility to write and erase the conducting paths at will within one and the same crystal or thin film.  

Among others, conducting DWs in the ferroelectric model system lithium niobate (\ce{LiNbO3}, LNO) have attracted concerted interests, since (a) their conductivity can exceed the corresponding bulk values by many orders of magnitude, (b) they are stable across a broad temperature range, and (c) they have already been well described in various previous works \cite{schroder_conducting_2012,schroder_nanoscale_2014,werner_large_2017,godau_enhancing_2017,kirbus_realtime_2019}. Here, \emph{fundamental aspects} such as the inherent relationship between the DW's geometrical inclination and the resulting electrical conduction, the role of the contact material, the typically non-ohmic nature of the respective current-voltage (IU) characteristics, or signatures for the thermally-activated behavior of DW electrical transport have been reported for selected samples, however, for a rather narrow temperature range so far \cite{schroder_conducting_2012,shur_timedependent_2013,werner_large_2017,chai_conductions_2021,geng_conductive_2021}. In parallel, there is a plethora of very recent \emph{application-related} results already demonstrating single \emph{electronic DW-based functionality} in either LNO single crystals or thin films, ranging from simple rectifying junctions 
\cite{werner_large_2017,zhang_erasable_2021,qian_domainwall_2022,suna_tuning_2023} towards more complex logic gates \cite{sun_in-memory_2022,suna_tuning_2023}, memristors \cite{chaudhary_lowvoltage_2020,kampfe_tunable_2020}, or transistors \cite{mcconville_ferroelectric_2020,chai_nonvolatile_2020,sun_in-memory_2022}, to name a few. 

Nevertheless, there are at least two crucial preconditions to meet for proper operation of any reliable \ce{LiNbO3} based DW device. First, the related device-specific parameters within an appropriate equivalent circuit have to be quantified, including the evaluation of both their general reproducibility and their temperature-dependence. Second, an in-depth understanding and modeling of the underlying electronic transport mechanism, which is not addressed comprehensively so far, has to be achieved. This is exactly the starting point of this work. In order to meet the first aspect, we present room-temperature current-voltage characteristics of a set of four virtually identically prepared DWs in single crystalline \SI{5}{\mol\percent\ \ce{MgO}}-doped \ce{LiNbO3}, postulate an equivalent-circuit model consisting of a parallel connection of two resistor/diode pairs (the R2D2 model), extract the corresponding resistances, saturation currents, as well as ideality factors, and discuss their asymmetry relative to the crystal orientation. For dealing with the second aspect, we analyze the temperature-dependent IU characteristics $I(U,T)$ of two exemplary DWs in detail, which allows us to extract the activation energies for the semiconductor-like intrinsic DW transport, and the effective Schottky barrier heights of the electrode-DW junction diodes. 

Finally, we further test the general applicability of our R2D2 model by analyzing the domain wall conductivity (DWC) data not only to z-cut bulk LNO single crystals, but equally to DWC observed in z-cut thin-film LNO (TFLN) and literature data on DWC in x-cut LNO. 

\section{Experimental}

\subsection{Preparation of \texorpdfstring{\ce{LiNbO3}}{LiNbO3} domain walls with enhanced electrical conductivity}
\label{sec:preparation}

For the present comparative study, four samples were cut from a commercial monodomain, \SI{5}{\mol\percent\ \ce{MgO}}-doped, congruent, \SI{200}{\micro\meter}-thick, z-cut \ce{LiNbO3} wafer, purchased from \textit{Yamaju Ceramics Co., Ltd.}, polished to optical quality. These crystal pieces measure \(5 \times \SI{6}{\milli\meter\squared}\) along their crystallographic x- and y-axis, respectively. In the following, these four samples are labeled \textit{DW-01} to \textit{DW-04}. Realizing the protocols described in detail earlier \cite{godau_enhancing_2017,godau_herstellung_2018} and in the SI-sec.~A, one single hexagonally-shaped reversely polarized domain was grown by laser-assisted poling within each sample, imaged by polarization-sensitive optical microscopy [see, e.g., SI-fig.~S1(b)], and -- after vapor-deposition of macroscopic Cr electrodes onto both crystal surfaces covering the DWs completely [cf.~\cref{fig:equivalent_circuit:sketch}] -- electrically tested by acquiring \SI{\pm 10}{V} standard current-voltage characteristics, which revealed a very low, nearly bulk-like conductivity with currents in the \SI{0.1}{pA}-range [SI-fig.~S1(c)].

\begin{figure}
    \centering
    \phantomsubfloat{\label{fig:DWC_stabilization:IV}}
    \phantomsubfloat{\label{fig:DWC_stabilization:time_dep}}
    \includegraphics[width=0.5\textwidth]{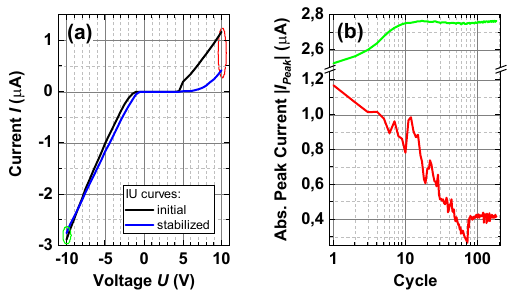}
    \caption{Current-voltage characteristics and current stabilization subsequent of the \emph{conductivity enhancement procedure} according to \citet{godau_enhancing_2017}, shown for sample \textit{DW-03}. (a) First (black) and last (blue) IU cycle obtained directly after the conductivity enhancement procedure and \SI{9}{h} later. (b) Evolution of the absolute value of the maximum current at \SI{+10}{V} (red) and \SI{-10}{V} (green) as a function of the number of measurement cycles. The IU~cycles were acquired between \SI{-10}{V} and \SI{+10}{V}, setting the measuring voltage in steps of \(\Delta U = \SI{0.5}{V}\) with time intervals of \SI{2}{s}.}
    \label{fig:DWC_stabilization}
\end{figure}

Subsequently, the DW conductivity was \emph{enhanced} by ramping up a high voltage [SI-fig.~S1(a)], provided by the voltage source of a \textit{Keithley 6517B} electrometer, while simultaneously monitoring the current flow [SI-fig.~S1(d)] according to Godau \textit{et al.}~\cite{godau_enhancing_2017}. As a result, the resistance of the DWs decreased significantly by up to seven orders of magnitude, as shown exemplarily again for sample \textit{DW-03} in \cref{fig:DWC_stabilization:IV} by the respective IU characteristics; corresponding data sets for all four samples were recorded as well and will be discussed later in \cref{sec:DWparameters} in detail. A stabilization process of the conductivity towards its final magnitude is observed on the time scale of several hours and shown in \cref{fig:DWC_stabilization:time_dep}. Thereafter, the conductivity was re-checked and proven to be time independent for at least one month. The exact measurement parameters (voltage sweep velocity, voltage increments, metallic measurement box) were the same as for acquiring the \emph{as-grown} IU curves described in the SI-sec.~A~2. 

Additionally, the 3-dimensional internal DW structure of two of the samples (\textit{DW-02} and \textit{DW-04}) were imaged by Cherenkov second-harmonic generation (CSHG) microscopy \cite{kampfe_optical_2014,kampfe_realtime_2015}, which clearly verified the known relationship between DW inclination and enhanced conduction. For experimental details and images we refer to SI-sec.~B and SI-fig.~S2, the latter showing a decisively altered, i.e., shrinked and inclined domain wall shape for the case of \textit{DW-02} and a strong structural change for the case of \textit{DW-04}, which we refer to as "domain explosion", as described and discussed by Kirbus \textit{et al.}~\cite{kirbus_realtime_2019} earlier.

\subsection{Quantitative analysis of room temperature IU-characteristics: the R2D2 model}
\label{sec:quant_analysis_RT_IU}

\begin{figure}
    \centering
    \phantomsubfloat{\label{fig:equivalent_circuit:sketch}}
    \phantomsubfloat{\label{fig:equivalent_circuit:circuit}}
    \includegraphics[width=0.5\textwidth]{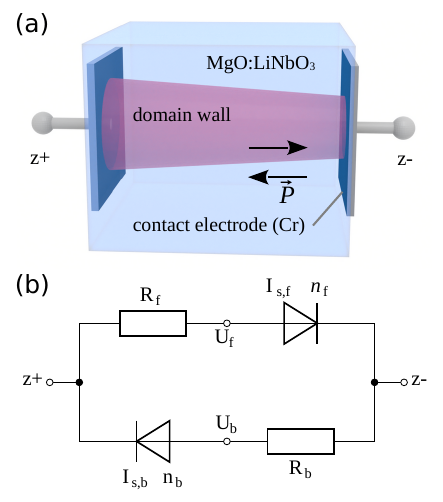}
    \caption{(a) Sketch of the sample configuration: \ce{LiNbO3} crystal with an inclined and thus conducting domain wall structure between two Cr electrodes at the crystal's z+ and z- surfaces. (b) Proposed R2D2 equivalent circuit consisting of a parallel connection of two diode-resistor combinations, which describes the IU curves of \ce{LiNbO3} domain walls contacted by Cr electrodes on both (z+/z-) crystal sides. The circuit elements can be quantitatively characterized by a non-linear nodal analysis at the intermediate potential nodes $U_f$ and $U_b$ based on Kirchhoff's law \cite{lipiansky_electrical_2013} in the way that the resistances $R_f$ and $R_b$ of the resistors as well as saturation currents ($I_{s,f}$, $I_{s,b}$) and ideality factors ($n_{s,f}$, $n_{s,b}$) of the two diodes are extracted from curve fitting procedures based on \cref{equ:kirchhoff,equ:shockley}.}
    \label{fig:equivalent_circuit}
\end{figure}

Since the typical \emph{post-enhancement} IU characteristics of a \ce{LiNbO3} domain wall with its two Cr~electrodes exhibits the shape as shown in \cref{fig:DWC_stabilization:IV}, including obviously (i) non-ohmic, diode-like regions for low voltages, but (ii) linear behavior for higher measurement voltages, with (iii) an additional clear asymmetry with respect to the voltage polarity, we heuristically postulate a \emph{parallel connection of two diode-resistor pairs}, sketched in \cref{fig:equivalent_circuit}(b), as the related equivalent circuit, where one pathway describes the "forward" and one the "backward" behavior along the DW. The four circuit elements are characterized by their resistances \(R\), saturation currents \(I_s\), and ideality factors \(n\), each in forward and backward direction (symbolized by the indices $f$ and $b$), respectively. To calculate the electric current through the circuit according to Kirchhoff's current law \cite{lipiansky_electrical_2013}, the currents at the nodes with the intermediate potentials \(U_f\) and \(U_b\) are considered. Due to charge conservation the currents flowing through the respective resistors and diodes must be equal at these nodes. Formally expressed, there is at least one voltage value \(U_f \in [U_{z-}, U_{z+}]\), for which the following relation holds:

\begin{equation}
    I_{resistor} = \frac{U_{z+} - U_f}{R_f}
        = I_{diode}(I_{s, f}; n_f; U_f - U_{z-}).
    \label{equ:kirchhoff}
\end{equation}

In case of a resistor and a diode that both have a monotonous IU characteristic, it is exactly one voltage value that exists.
Thereby \(I_{diode}\) is represented by the well-established Shockley equation \cite{rhoderick_metalsemiconductor_1988}: 

\begin{equation}
I_{diode}(I_s; n; U) = I_s \left[ \exp \left(
        \frac{U}{n k_B T} \right) - 1 \right],
\label{equ:shockley}
\end{equation}
with \(k_B\) being the Boltzmann constant, and \(T\) the absolute temperature. This choice of \(U_f\) ensures that no charges accumulate at the intermediate node, and that the current flow is time-independent. For \(U_b\) and the circuit elements in \emph{backward} direction, analogous considerations are taken into account. 

Thus the 2-resistors/2-diodes (R2D2)-model exhibits six free parameters: $R_{f}$; $R_{b}$; $I_{s,f}$; $I_{s,b}$; $n_{f}$; $n_{b}$, which can be fitted by numerical treatment of \cref{equ:kirchhoff}. Since the characteristic IU~curves were defined by more than \num{40} experimentally obtained measurement points (cf.~SI-sec.~A~2 and \cref{sec:preparation}), the convergence of the optimization process
is granted. Before applying the fitting routine, parameters were manually adjusted to the right order of magnitude, to ensure convergence. To take account for the possibly broad intervals for the fitting parameters, the logarithm to the base 10 of the parameters' values was optimized instead of the parameters themselves. The optimization was performed using a trust region reflective algorithm \cite{branch_subspace_1999} with least-squares cost function, as implemented in the \textit{python3} library \textit{scipy} \cite{johansson_numerical_2019}. 

\subsection{Investigation of the electrical-transport mechanism by temperature-dependent IU-curves}
\label{sec:mat_meth_temp_dep}

In order to (i) figure out the precise electrical transport mechanism through the DW, (ii) to derive the corresponding characteristic parameters such as the activation energy or the barrier height, and (iii) to check for the temperature stability of the circuit parameters in general, temperature dependent IU measurements from \SI{320}{K} down to around \SI{80}{K} were performed with two of the four samples, i.e., \textit{DW-01} and \textit{DW-04}. A liquid nitrogen bath cryostat (\textit{Optistat DN} by \textit{Oxford Instruments}) was used, controlling the temperature by two independent Pt-100 platinum resistance sensors, one positioned at the heat exchanger and the other directly next to the sample. The operation of the cryostat, comprising gas flow regulation, heating control, and temperature reading at the heat exchanger, were accomplished via an \textit{ITC 503} temperature controller by \textit{Oxford Instruments}, while the Pt-100 sensor near the sample was read out by a \textit{Keithley 196} digital multimeter. Full IU characteristics in the \SI{\pm 10}{V} range were acquired for 40 different logarithmically distributed temperatures with a \textit{Keithley 6517B} electrometer in steps of \(\Delta U = \SI{0.5}{V}\) with \(dU/dt = \SI{0.5}{V/s}\) in two-point geometry with wires shielded up to the probe head. The temperature setpoints were changed stepwise and three IU cycles were recorded, however, always after reaching thermal equilibrium, i.e., after a waiting period of around \SI{30}{minutes} when having set a new setpoint temperature. To eliminate transient effects, all subsequent calculations were performed with the third IU cycle only. While changing the temperature, the two electrodes were short-circuited via the electrometer to achieve an equalization of pyroelectrically generated charges. Furthermore, spurious temperature fluctuation during IU measurements were proven to be less than \SI{0.01}{K}. Thus pyroelectric effects are neglected in all following evaluations.

\begin{figure}
    \centering
    \includegraphics[width=0.45\textwidth]{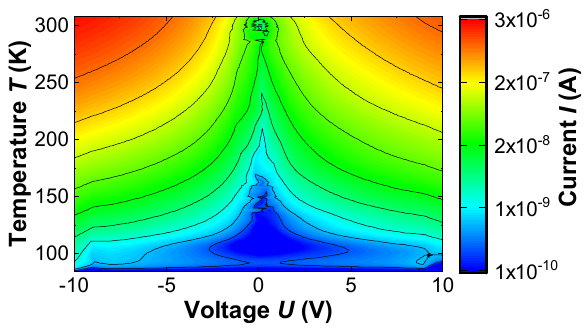}
    \caption{Temperature-dependent current-voltage data, depicted as a \emph{heat map} heat map with $40 \times 40$~measured current values as a function of both measuring voltage and temperature (data from \emph{DW-04}).}
    \label{fig:current-heatmap}
\end{figure}

In sum, a "3D" data field $I(U,T)$ with current values $I$ measured at $40 \times 40$ voltage-temperature combinations $(U,T)$ was collected (\cref{fig:current-heatmap}). The IU characteristics at fixed temperature were evaluated analogously to the processing of the room temperature curves described in \cref{sec:quant_analysis_RT_IU}. As a result, the temperature dependences of DW resistances [$R_{f}(T); R_{b}(T)$], diode saturation currents [$I_{s,f}(T); I_{s,b}(T)$], and ideality factors [$n_{f}(T); n_{b}(T)$] could be established. 
First, the $R(T)$ characteristics were brought to the form of Arrhenius plots $[\ln(R)(1/T)]$, which allowed us extracting the activation energy $E_a$. In order to decide to which precise $R(T)$ curve form the data should be fit to extract $E_a$, a preliminary, tentative evaluation step was carried out. Thereby, a number of electrical-transport models, such as thermally-activated hopping and different polaron hopping and variable-range hopping models, were tested for the exemplary case of sample \textit{DW-01} with the results that \textit{simple thermally-activated hopping} transport with the following temperature dependence of the conductivity $\sigma$ (being equivalent to $R$):

\begin{equation}
    \sigma(T) = \tilde{\sigma}_0 \exp \left( -\frac{E_a}{k_B T}\right),
    \label{equ:sigma_general}
\end{equation}
where \(\tilde{\sigma}_0\) symbolizes a constant prefactor related to the sample geometry, appears to be the most probable process here, which fully agrees with assumptions used by other authors before \cite{werner_large_2017, kampfe_tunable_2020}. This allows for a linear fitting of the Arrhenius plots with $-E_a/k_B$ being the slope. The full analysis including a listing and a short explanation of all considered models is found in SI-sec.~C (based on the more detailed work of ref.~\cite{zahn_nonlinear_2022}, taking into account a number fundamental works on transport phenomena, i.e., refs.~\cite{hoffmann_how_1987,baranovski_charge_2006,miller_impurity_1960,mott_conduction_1968,efros_coulomb_1975,mott_electronic_2012,laiho_mechanisms_2005}).

Second, the curves for the saturation currents $I_s(T)$ were fitted using the theoretical relationship derived from the thermionic emission model, which, in brief, describes the transport across an energy barrier as a combination of thermal activation and tunneling through the latter on the activated energy level (see, e.g.,~\citet{rhoderick_metalsemiconductor_1988}):

\begin{equation}
	I_s = A^\star T^2 \exp \left( \frac{- q \Phi_{eff}}{k_B T} \right),
	\label{equ:theory:schottky_is_rhoderick}
\end{equation}
with \(\Phi_{eff}\) being the effective potential barrier height, and \(A^\star\) a material-specific parameter, known as the Richardson constant. Thus, this evaluation supplied us with estimates for the effective Schottky barrier heights of the DW-metal contacts.

\section{Results and discussion}

\subsection{Quantifying resistor and diode parameters from room-temperature IU characteristics}
\label{sec:DWparameters}

\begin{figure}
    \centering
    \includegraphics[width=0.45\textwidth]{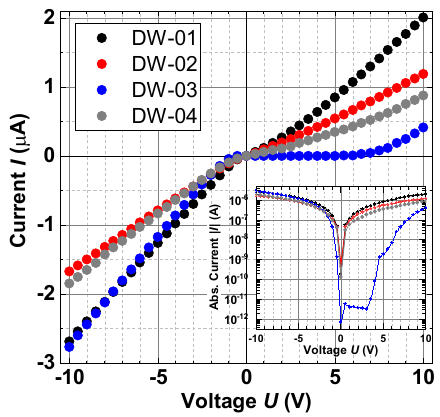}
    \caption{Room-temperature current-voltage characteristics of a set of four domain walls in single-crystal \ce{LiNbO3}, contacted macroscopically with vapor-deposited \ce{Cr} electrodes, all reproducibly revealing a non-ohmic and asymmetric DWC. The inset depicts the corresponding semilogarithmic plot (see also SI-fig.~S3 for a close-up view). For both the geometric interpretation and the assigned R2D2 equivalent circuit, which combine ohmic and diode-like character in a single concept, see \cref{fig:equivalent_circuit}(b) again.}
    \label{fig:dc_rt-ivcurves}
\end{figure}

As one key result, \cref{fig:dc_rt-ivcurves} comparatively displays the IU characteristics of the four virtually identically prepared conductive domain walls in bulk LNO at room-temperature. All curves show clear non-ohmic behavior for low measuring voltages and a rather linear progression for larger applied voltages, with a clear asymmetry towards the polarity of the measuring voltage. As indicated in \cref{sec:quant_analysis_RT_IU}, we fit all four curves according to the R2D2 double-diode/double-resistor equivalent circuit model (\cref{fig:equivalent_circuit}(b), \cref{sec:quant_analysis_RT_IU}) and obtain the numerical values for the six free parameters as summarized in \cref{tab:dc-parameter} (cf.~SI-table~S2 for uncertainties), which we discuss more closely in the following.

\begin{itemize}
    \item \textbf{Resistances:} Typical values between \num{2.6} and \SI{7.2}{\mega\ohm}, which lie all within the same order of magnitude, are observed for \(R_f\) and \(R_b\). This also means a conductivity enhancement of 7--8 orders of magnitude as compared to the as-grown domain walls before application of the "enhancement" protocol [cf.~SI-fig.~S1(c)], which is the expected and desired result due to the enlarged domain wall inclination with respect to the crystal's z-axis.
    
    \item \textbf{Saturation currents:} Here, a nominally large variation over five orders of magnitude between \SI{10}{\pico\ampere} and \SI{1}{\micro\ampere} turns out at first glance. However, when excluding the $I_{s,f}$ value for \emph{DW-03} which is probably caused by a peculiarity in the real structure of the electrode-DW junction as clearly seen from the logarithmic current plot in the inset of \cref{fig:dc_rt-ivcurves} and in SI-fig.~S3, the saturation currents cover only two orders of magnitude.
    
    \item \textbf{Ideality factors:} The values for \(n_f\), \(n_b\) are much larger than two, indicating significant differences between conductance along domain walls and conventional semiconductors (with the latter having typical $n$ values between \num{0.5} and \num{2}). However, this phenomenon of anomalously high $n$ values also occurs for highly-doped semiconductors (for silicon above \(N_d \approx \SI{e19}{\per\cubic\meter}\) at \SI{300}{\kelvin}) and is described by the field emission case of the thermionic-emission theory \cite{rhoderick_metalsemiconductor_1988}. Transferred to the domain wall, it indicates a high hopping site density inside the domain wall that is in agreement with former theoretical calculations on \ce{LiNbO3} \cite{eliseev_static_2011}. Notably, the ideality factors in forward direction \(n_f\) are much larger as compared to the backward direction \(n_b\). 
\end{itemize}

There are natural reasons for the inequality of forward and backward parameters, namely, first, the geometrical DW asymmetry between z$+$ and z- side due to the domain growth process starting at the z$+$ towards the z- side. Second, on an even more fundamental level, the general intrinsic asymmetry of the two different \ce{LiNbO3} surfaces is reflected in their different surface terminations, the different processes of charge compensation, and the subsequent dramatically different ionization energies (6.2~eV vs. 4.9~eV) and thus work functions, as shown experimentally by photoelectron spectroscopy in the past \cite{yang_polarization_2004} and supported by theoretical calculations as well \cite{holscher_adsorption_2012,holscher_modeling_2014,sanna_temperature_2014}. Consequently, it appears to be logical that these two very different crystal surfaces form clearly distinguishable junctions with the Cr electrodes with side-specific (though not known in detail) electronic-defect state distributions and thus band alignments, which are finally visible as direction-dependent equivalent-circuit parameters.   

The rather high coefficient of determination \(R^2\) achieved for all samples indicates that the R2D2 equivalent-circuit model describes the conduction parameters adequately well.

In addition to the two considered current paths, two further channels may contribute to the overall conductance, one having two diodes in opposite direction and one consisting of a single resistor only. While the first path can only weakly conduct due to the reverse-biased diode, the second path would exhibit a purely ohmic IU characteristic, as was observed by \citet{werner_large_2017} and \citet{godau_enhancing_2017}. Both turned out to be of minor influence in our experiments but can not be excluded in general.

\begin{table}
    \center
    \begin{tabular}{|l|r|r|r|r|r|r|r|}  \hline
		\thead{sample} & \thead{\(R_f\)} & \thead{\(I_{s, f}\) }
		 & \thead{\(n_f\)} & \thead{\(R_b\)} & \thead{\(I_{s, b}\)}
		 & \thead{\(n_b\)} & \(R^2\) \\ 
		 & \thead{[\si{\mega\ohm}]} & \thead{[pA]}
		 &  & \thead{[\si{\mega\ohm}]} & \thead{[pA]}
		 &  &  \\ 
		 \hline
		\textit{DW-01} & 3.68 & \num{1.23e5} & 35.7 & 3.44 & 430 & 5.33 & 0.990 \\
		\textit{DW-02} & 7.16 & \num{9.83e4} & 23.6 & 5.51 & 100 & 5.25 & 0.980 \\ 
		\textit{DW-03} & 2.84 & 12.8 & 33.6 & 3.07 & 211 & 6.41 & 0.951 \\
		\textit{DW-04} & 2.59 & \num{3.41e5} & 240 & 4.48 & 4913 & 19.8 & 0.957 \\
		 \hline
	\end{tabular}
    \caption{Equivalent circuit parameters obtained by modeling the IU characteristics shown in \cref{fig:dc_rt-ivcurves} according to the R2D2 model [\cref{fig:equivalent_circuit}(b)] via a least-square fit based on Kirchhoff's current law. Note that $R^2$ in the last column denotes the coefficient of determination here. See also SI-table~S2 that additionally tabulates the uncertainties for all fit parameters.}
    \label{tab:dc-parameter}
\end{table}

One may speculate whether there is a fundamental reason that a \textit{parallel} connection of two current paths appears to be the most suited here. Such a reasoning is also motivated by previous results obtained by \citet{godau_enhancing_2017} and \citet{wolba_resistor_2018} on the nonuniform local distribution of the conductance. Thus, the current is bound to distinct channels (preferably along the domain wall corners) that are separated from each other.

Apart from the above more electrotechnical viewpoint focusing on the circuit-element quantification, we now proceed with the physical interpretation of the suggested circuit elements in the R2D2 model, shown in \cref{fig:equivalent_circuit}(b), distributed to the two separate current paths. Heuristically, the diodes represent the Schottky barriers between the metal electrodes and the DW, while the resistors reflect the intrinsic DWC. In the following section, we obtain two more characteristic parameters of the two transport contributions by temperature dependent IU measurements: (i) the activation energy $E_a$ for the transport along the DW, and (ii) the effective barrier height $\Phi_{eff}$ for the metal-DW junction.

\subsection{Analysis of the underlying carrier transport processes through temperature dependent DW~current measurements}
\label{sec:results_lowT}

\begin{figure}
    \centering
    \includegraphics[width=0.5\textwidth]{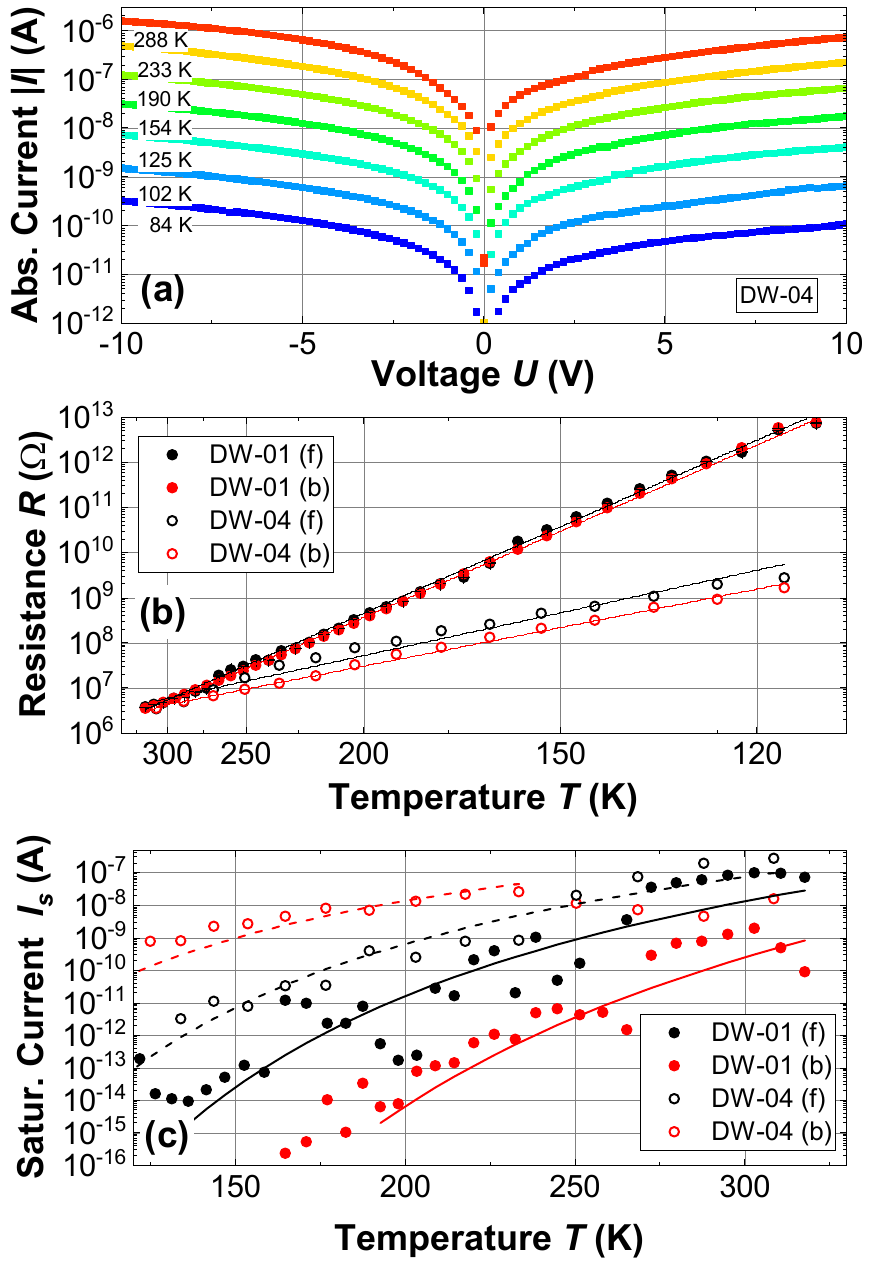}
    \phantomsubfloat{\label{fig:dc_lowT:ivcurves}}
    \phantomsubfloat{\label{fig:dc_lowT:resistance}}
    \phantomsubfloat{\label{fig:dc_lowT:saturationI}}
    \caption{(a) Temperature dependent current-voltage curves in logarithmic representation, exemplarily shown for \emph{DW-04}.
    Equivalent circuit parameters (b) $R$ (dots: experimental data, lines: Arhennius-law fits) and (c) $I_s$ [dots: experimental data, lines: fits according to the thermionic emission model, cf.~\cref{equ:theory:schottky_is_rhoderick}] as a function of temperature, which confirm the semiconductor-like intrinsic conductivity in LNO domain walls between \num{110} and \SI{320}{\kelvin}, providing estimates for the activation energies $E_a$ and the effective Schottky barrier heights $\Phi_{eff}$ via the respective fit parameters, see \cref{tab:parameters_lowt_diode_fitting}. The fits were performed over the respective temperature ranges indicated by the lines only. The fitting ranges had to be limited in this way, since we had to disregard data points of high relative uncertainty. Note the reciprocal scaling of the temperature axis in panel (b).}
    \label{fig:dc_lowT}
\end{figure}

To achieve an in-depth understanding of both the transport across the Schottky barrier at the two electrode-DW interfaces and along the DW itself, IU characteristics for samples \emph{DW-01} and \emph{DW-04} at different temperatures between \num{80} and \SI{320}{\kelvin} were acquired, as exemplarily shown for selected temperatures in \cref{fig:dc_lowT:ivcurves}. Obviously, the current level decreases with decreasing temperature, as is typical for a \emph{semiconducting} material, while the general shape of the IU characteristics does not significantly change with temperature, showing the same features as discussed in the previous section. Due to the latter fact, we extracted the $R$, $I_s$, $n$ values analogously to the room temperature parameters, but can plot them now as a function of temperature, as displayed in \cref{fig:dc_lowT:resistance,fig:dc_lowT:saturationI}, as well as in the SI-fig.~S4, respectively.

\begin{table}[ht]
    \centering
    \begin{tabular}{| l |
            >{\collectcell}r<{\endcollectcell}
			@{\({}\pm{}\)}
			>{\collectcell}l<{\endcollectcell} |
			>{\collectcell}r<{\endcollectcell}
			@{\({}\pm{}\)}
			>{\collectcell}l<{\endcollectcell} |
			>{\collectcell}r<{\endcollectcell}
			@{\({}\pm{}\)}
			>{\collectcell}l<{\endcollectcell} |
			} \hline
		\thead{sample} &
        \multicolumn{2}{c|}{\thead{\(E_a\) [eV]}} & \multicolumn{2}{c|}{\thead{\(A^\star\) [\unit{\nano\ampere\per\kelvin\squared}]}} & \multicolumn{2}{c|}{\thead{\(\Phi_{eff}\) [eV]}} \\ \hline
        DW-01 (f) & 0.2291 & 0.0010 & 775 & 1954 & 0.50 & 0.05 \\
        DW-01 (b) & 0.2290 & 0.0016 & 19 & 32 & 0.305 & 0.029 \\
        DW-04 (f) & 0.1008 & 0.0019 & 0.16 & 0.28 & 0.106 & 0.020 \\
        DW-04 (b) & 0.107 & 0.005 & 2.1 & 2.89 & 0.203 & 0.022 \\ \hline
    \end{tabular}
    \caption{Activation energy $E_a$, Richardson constant \(A^\star\), and Schottky barrier height $\Phi_{eff}$ tabulated for the two inspected bulk DWs in LNO, as derived from the curve fits of $R_{f,b}(T)$ and $I_{s;f,b}(T)$ in \cref{fig:dc_lowT:resistance,fig:dc_lowT:saturationI}.}
    \label{tab:parameters_lowt_diode_fitting}
\end{table}

First, the obtained resistances $R$ [\cref{fig:dc_lowT:resistance}] follow an Arrhenius-like temperature characteristics, as expected. The Arrhenius law is observed across the full temperature range, proving the stability of the intrinsic conduction process, which might be easy to account for in a potential DW nanoelectronic device. The numerically extracted activation energies $E_a$ (shown in \cref{tab:parameters_lowt_diode_fitting}) match well between forward and backward direction for each sample, but differ significantly between them. Though the exemplary analysis of of the $\sigma(T)$ dependence, as discussed in \cref{sec:mat_meth_temp_dep} and SI-sec.~C, it was not unambiguously possible to clarify whether we are faced with simple-thermal activation, adiabatic or non-adiabatic polaron hopping, there is circumstantial evidence, which all together point towards hopping of small free electron-polarons, as also observed in bulk LNO, being the dominant transport process along the DWs as well. First, recent in-situ-strain~\cite{singh_tuning_2022} and Hall-effect~\cite{beccard_hall_2023} experiments on similarly conductivity-"enhanced" DWs showed negative charge carriers to be the majority carriers, and, second, the numerical $E_a$ values between 0.101~eV and 0.229~eV derived here (table~\ref{tab:parameters_lowt_diode_fitting}) are in full accordance with reported activation energies for electron-polaron hopping in bulk LNO, as summarized, e.g., in the review article by Reichenbach \emph{et al.}~\cite{reichenbach_polaronmediated_2018}, where the (also rather large) interval between 0.1 and 0.24~eV is stated.

Second, the saturation currents $I_s$ of the diode component [\cref{fig:dc_lowT:saturationI}] can be satisfactorily fitted by \cref{equ:theory:schottky_is_rhoderick} reflecting the validity of the thermionic emission model for the electrode-DW Schottky contact. The effective Schottky barrier $\Phi_{eff}$ is estimated between \num{0.1} to \SI{0.5}{\electronvolt} (also listed in \cref{tab:parameters_lowt_diode_fitting}). A likely source of this rather large range are variations of the individual electronic-defect state distribution at the metal-domain wall interface introduced during the metal-electrode deposition and conductivity enhancement procedure, whereas the latter has the decisively larger impact~\cite{kis23}. An interpretation of the extract Richardson constants \(A^\star\), which vary over three orders of magnitude, is not easily possible, since they depend on several barely known quantities such as the relative electron mass \(m^\star\) and the barrier cross section width. In all four regarded cases, the uncertainty of \(A^\star\) seems to be heavily overestimated due to the exponential transformation, while \(\log_{10}(A^\star)\) is still a well defined quantity with a relative uncertainty of less than \SI{10}{\percent}. 

Third, the evaluation of the ideality factors $n$ shows a more ambiguous picture (see SI-fig.~S4). Based on the thermionic-emission theory, only a very weak temperature dependence is expected for the ideality factors due to changes of the effective electron mass \cite{rhoderick_metalsemiconductor_1988, rudan_physics_2018}. Furthermore, the ideality factor strongly depends on the effective hopping site density, which is independent of temperature. However, apart from the case of sample \textit{DW\nobreakdash-01} in backward direction, which shows indeed a rather constant value over the covered temperature range, the results for the remaining cases exhibit rather strong fluctuations between 15 and larger than 80, caused by the fragile position of $n$ within the argument of the exponential function. On the other hand, even the quite scattered data supports the trend towards $n$ values being considerably larger as compared to standard semiconductor diodes.

\begin{figure}
    \centering
    \includegraphics[width=0.45\textwidth]{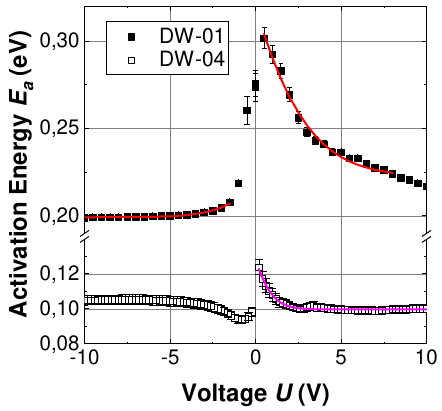}
    \caption{Activation energies, as derived from the currents' Arrhenius plots, as a function of the measuring voltage: dots reflect the experimental data, solid lines the fit curves according to \cref{equ:activation_energy}.}
    \label{fig:ea-vs-U}
\end{figure}

As an interesting and illustrative side note, the acquired 3D-data set $I(U,T)$ -- plotted as a heat map in \cref{fig:current-heatmap} -- allows extracting the activation energy $E_a$ directly as the slope from the $\ln(I)(1/T)$ current-vs.-temperature curves. These values are shown in \cref{fig:ea-vs-U}, plotted as a function of measuring voltage $U$, together with the partial fits to the theoretical $E_a(U)$ dependence. Based on \cref{equ:sigma_general}, the latter is obtained by calculating the partial derivative of $\ln I$ with respect to $1/T$, as worked out in detail in SI-sec.~G, using \cref{equ:shockley,equ:theory:schottky_is_rhoderick}, finally resulting in:

\begin{equation}
		E_a := - k_B \frac{\partial \ln I}{\partial 1/T} 
		= E_0 - A \cdot \frac{U/U_c}{1 - \exp(-U/U_c)}, 
	\label{equ:activation_energy}
\end{equation}
with the fit parameters $E_0$, $A$, and $U_c$. However, the qualitative agreement of the experimental $E_a(U)$ curve with the (partial) fits according to \cref{equ:activation_energy} is convincing for both samples. The curves show a characteristic strong increase of $E_a$ at low voltages, which clearly supports our central assumption that at low fields the barrier at the electrode-DW junction dominates the transport behavior of the electrode-DW system. The significant difference of the constant activation energy at large electric fields between 0.10 and 0.22~eV for the two tested DWs is astonishing at first glance, but is in full agreement with the activation energies derived from the Arrhenius plots of the resistances before, which show nearly the same rather different values for the two inspected samples. We refrain from discussing the fit parameters $E_0$, $A$, and $U_c$ in detail, since they refer to a kind of an effective temperature and thus their strict physical meaning is not straightforward to determine.

\subsection{Applying the R2D2 model to domain wall conductance in thin-film \texorpdfstring{\ce{LiNbO3}}{LiNbO3}}
\label{sec:results_TFLN}

\begin{figure}
    \centering
    \includegraphics[width=0.45\textwidth]{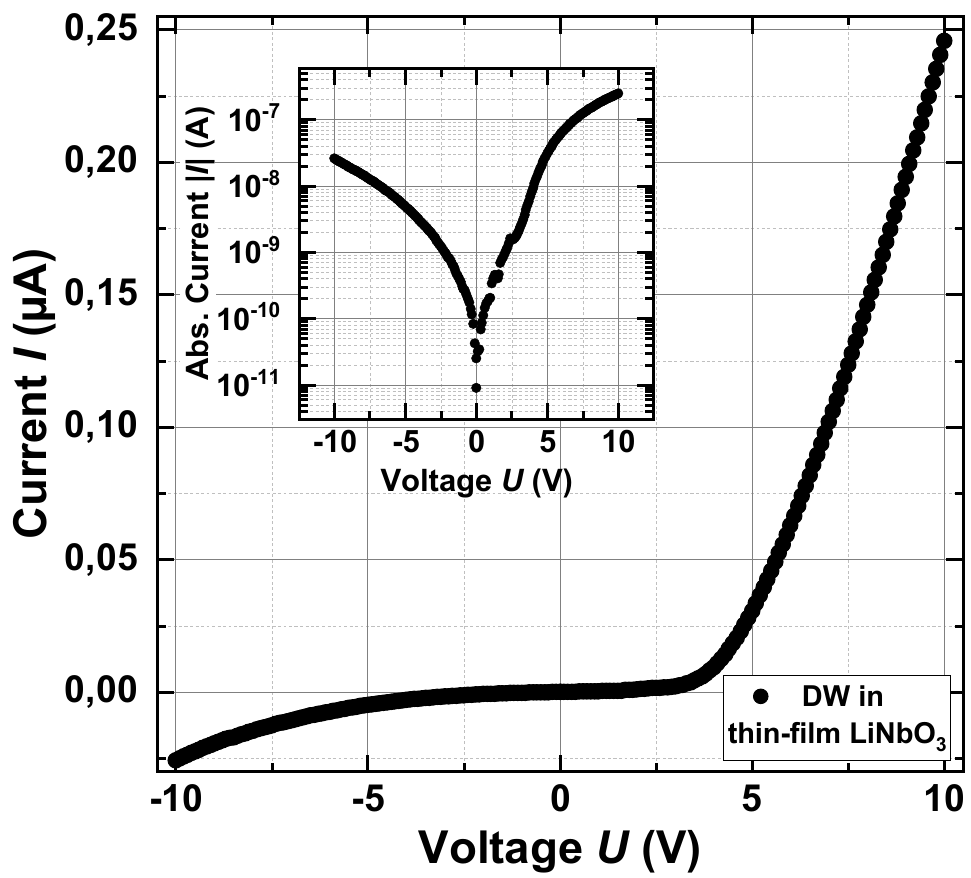}
    \caption{Current-voltage characteristics of a thin-film \ce{LiNbO3} domain wall induced by an AFM tip, contacted with \ce{Ag} electrodes, revealing a non-ohmic and asymmetric current, depicted as linear and semilogarithmic (inset) plot.}
    \label{fig:dc_TFLN_ivcurves}
\end{figure}

To test whether or not our R2D2 model is of general use for interpreting domain wall conductance (DWC), we analyze below the DWC from two distinctly different samples:

\begin{itemize}
    \item DW arrays that have been written into a z-cut, 500~nm thin TFLN sample using a larger bias voltage applied to the tip of a scanning force microscope (for details see SI-sec.~H and ref.~\citep{mccluskey_ultrahigh_2022}). We processed these current-voltage characteristics of the DW array shown in \cref{fig:dc_TFLN_ivcurves} in the same way as accomplished for the IU curves in \cref{fig:dc_rt-ivcurves}. The curve fitting (for a summary of all results refer to SI-table~S3) succeeded with an excellent $R^2$ value of \num{0.99995}, resulting in resistances being a factor of ten larger than observed for the single-crystal DWCs, i.e., $R_f = \SI{18}{M\ohm}$ and $R_b = \SI{94}{M\ohm}$, and diode saturation currents being smaller on trend, namely $I_{s,f} = \SI{16}{pA}$ and $I_{s,b} = \SI{507}{pA}$. The extracted values of the forward and backward diode ideality factors, which are $n_f = 23$ and $n_b = 76$, surprised in comparison to the single-crystal results, where $n_f$ appears to be systematically larger than $n_b$. Thus, the data might indicate a different mechanism responsible for the directional asymmetry in the thin-film sample, which is supported by the findings by \citet{suna_tuning_2023} showing that near surface domain wall bending results in a significant contribution to the diode like response. Nevertheless, an in-depth clarification needs a more systematic approach, which is out of scope for the study here.

    \item We applied the R2D2 model to literature DWC data that were recorded at DWs in an x-cut TFLN sample by \citet{qian_domainwall_2022}. The analysis of this data results in resistances in the G$\Omega$~range, $I_s$ values in the pA~range, and similarly high ideality factors as above, all of them with satisfying $R^2$ values as well (see SI-table~S4 for the numerical values).
\end{itemize}

\section{Summary and Outlook}

In this study, ferroelectric conductive domain walls (CDWs) were engineered into \SI{200}{\micro\meter}-thick \SI{5}{\mol\percent\ \ce{MgO}}-doped \ce{LiNbO3} single crystals and contacted by macroscopic vapor-deposited chromium electrodes at both crystal sides. Current-voltage (IU) characteristics in the \SI{\pm 10}{V} range were recorded comparatively at a set of four such CDWs, which exhibited reproducibly asymmetric non-ohmic characteristics. Thus, an equivalent-circuit model, the R2D2 model, consisting of a parallel connection of two resistor/diode pairs was postulated empirically, which allowed us to fit the IU curves using Kirchoff's current law together with Shockley's diode equation, ending up in a systematic quantification of typical resistance ranges and diode parameters (saturation current, ideality factor) for this specific DW-electrode configuration in forward and backward direction, which indeed showed systematically different values due to the intrinsically unequal crystallographic and electrochemical behavior of the z$+$ and z$-$ LiNbO$_3$ surfaces. The model was also successfully applied to exemplary IU characteristics of differently created DWs in thin-film lithium niobate and might be generally usable within a standardized analysis routine of domain-wall related IU characteristics in the future.

From additional temperature dependent IU recordings at two selected CDWs, we empirically assigned (i) the diodic (nonlinear) part around zero measuring voltage to the influence of the CDW-electrode junction showing thermionic emission in the vicinity of a Schottky barrier, and (ii) the ohmic (linear) part at higher bias voltages to the intrinsic conduction within the domain wall. The latter was further identified to behave thermally-activated semiconductor-like, with activation energies between 100 and 230~meV, which quantitatively match to free electron-polaron hopping, as derived from the slope of the linear Arrhenius plots of resistances. Finally the effective Schottky barrier heights of the DW-electrode junctions were derived from the temperature dependence of the diode saturation currents.

Our results raise a number of questions to be addressed in the future. First, the microscopic nature of the electric-current paths along the CDWs was not completely clarified due to the usage of macroscopic electrodes. Here, a scanning probe microscopy based complementary investigation, especially employing conductive atomic force microscopy to directly contact different regions of the domain wall by the tip and capture local IU characteristics, are needed, which could potentially lead to a more generalized equivalent circuit model. Second, from a statistical point of view, an investigation of a decisively broader set of CDWs including IU and CSHG microscopy data of all specimen would allow us to correlate all relevant DW fabrication parameters to the final electrical performance in terms of the equivalent circuit parameters and to substantiate functional structure-property relationships. After having disentangled two different conduction contributions, a third future challenge is the control and optimization of the electrode-DW junction by varying the contact metal on the one hand and by a higher degree of automatization during the preparation process on the other hand.

\section*{Acknowledgements}

We acknowledge financial support by the Deutsche Forschungsgemeinschaft (DFG) through the CRC~1415 (ID: 417590517), the FOR~5044 (ID: 426703838; \url{https://www.for5044.de}), as well as through the Dresden-W\"urzburg Cluster of Excellence on "Complexity and Topology in Quantum Matter" - ct.qmat (EXC~2147, ID: 39085490). This work was supported by the Light Microscopy Facility, a Core Facility of  the CMCB Technology Platform at TU Dresden. M.Z. acknowledges funding from the Deutsche Forschungsgemeinschaft via the Transregional Collaborative Research Center TRR~360, the German Academic Exchange Service via a Research Grant for Doctoral Students (ID: 91849816), the Studienstiftung des Deutschen Volkes via a Doctoral Grant and the State of Bavaria via a Marianne-Plehn scholarship. I.K.'s contribution to this project is also co-funded by the European Union and co-financed from tax revenues on the basis of the budget adopted by the Saxon State Parliament.


\bibliography{main}

\newpage

\setcounter{table}{0}
\renewcommand{\thetable}{S\arabic{table}}  

\setcounter{page}{1}
\renewcommand{\thepage}{S\arabic{page}} 

\setcounter{equation}{0}
\renewcommand{\theequation}{S.\arabic{equation}} 

\setcounter{figure}{0}
\renewcommand{\thefigure}{S\arabic{figure}}

\setcounter{section}{0}
\renewcommand{\thesection}{\Alph{section}}

\renewcommand{\thesubsection}{\arabic{subsection}}

\onecolumngrid


\section*{Supplementary Information}

\section{More details on domain wall preparation in single crystal lithium niobate}
\label{sec:app:preparation}

\subsection{Writing of domains by UV-assisted poling with liquid electrodes}
\label{sec:app:poling}

\begin{figure}[h!]
    \centering
    \includegraphics[width=0.65\textwidth]{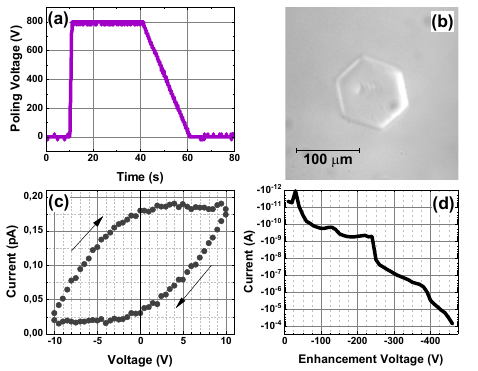}
    \phantomsubfloat{\label{fig:DW_prep:pulse}}
    \phantomsubfloat{\label{fig:DW_prep:image}}
    \phantomsubfloat{\label{fig:DW_prep:IV}}
    \phantomsubfloat{\label{fig:DW_prep:enhancement}}
    \caption{Engineering of conductive ferroelectric domain walls in a \ce{LiNbO3} single crystal sample (pictures refer to \textit{DW-03}, as an example) -- key features of selected preparation steps within the course of the protocol described by Godau \textit{et al.}~[13]: \textbf{(a)} Application of a high-voltage pulse (\SI{4}{kV/mm}) to a monodomain \ce{LiNbO3} crystal via liquid-electrodes under UV illumination at \SI{325}{nm} in order to nucleate a single domain of opposite polarization. \textbf{(b)} Result of the poling process -- a hexagonal domain imaged by polarization-sensitive optical microscopy; for a 3-dimensional visualization refer to SI-\cref{fig:cshg}. \textbf{(c)} Current-voltage curve of the \emph{as-grown} DW reflecting mainly the charging process of a parallel plate capacitor formed by the Cr electrodes on both crystal sides and the DW, indicating a high DW resistivity in the \si{\tera\ohm} range. \textbf{(d)} Application of a large voltage opposite to the poling direction in order to enhance the domain wall conductivity with simultaneous current monitoring. The total time of the "enhancement" voltage ramp counts \SI{75}{s}.}
    \label{fig:DW_prep}
\end{figure}

In a first preparation step, one single, typically hexagon-shaped, ferroelectric domain was created within each LNO sample by a laser-assisted poling procedure employing liquid electrodes as described in more detail in refs.~[13,26]. Therefore a voltage pulse of \SI{800}{V} height and \SI{40}{s} duration as depicted in \cref{fig:DW_prep:pulse} was applied between the two z-faces of the crystal. The result of the poling procedure -- a domain of around \SI{100}{\micro\meter} diameter -- was monitored by polarization-sensitive optical microscopy, see \cref{fig:DW_prep:image}.

\subsection{Conductivity measurements of as-grown domain walls}
\label{sec:app:cond_as_grown_DWs}

In this as-grown state the domain wall typically shows a comparably high resistance in the range of \SI{e13}{\ohm} as visible in the current-voltage (IU) characteristics in \cref{fig:DW_prep:IV} recorded with a \textit{Keithley 6517B} electrometer after having covered the domain on both crystal z-faces with vapor-deposited Cr electrodes of \SI{10}{nm} thickness and \SI{1}{mm^2} lateral area. In principle, the behavior does not differ from the bulk. The IU curves at room temperature were recorded by multiple voltage sweeps between \SI{-10}{V} and \SI{+10}{V} and vice versa, resulting in closed cycles with voltage increments of \SI{0.5}{V} per \SI{2}{s}, i.e., 40 data points for one "half cycle". The voltage is denoted with respect to z$+$ while keeping z- grounded. During these measurements the samples were positioned in a metal box using coaxial wiring to reduce both electrically and mechanically induced noise. A typical \emph{as-grown} IU curve, as shown for the third recorded cycle (to avoid initial transient effects) for \textit{DW-03} in \cref{fig:DW_prep:IV}, appears to be point symmetric, however, with a clear positive offset current of around 0.10~pA, which is an electrometer-induced artefact. Apart from that, the IU curve basically reflects the charging process of the parallel plate capacitor formed by the two electrodes with the \ce{LiNbO3} crystal as dielectric. Assuming a voltage sweep with a constant ramp rate \(dU/dt\), the charging current \(I_{charge}\) can be derived from the well-known formula describing the capacitance \(C\) of a parallel plate capacitor:

\begin{equation}
    C=\epsilon_0 \epsilon_r \frac{A}{d} = \frac{\delta Q}{\delta U} = \frac{dQ/dt}{dU/dt} = \frac{I_{charge}}{dU/dt} \quad.
\end{equation}

Inserting relevant values, i.e., the dielectric constant \(\epsilon_r = 33\) (of MgO-doped LiNbO$_3$), a capacitor plate area of \(A=\SI{1}{mm^2}\) (Cr contact size), a thickness of the dielectric of \(d~=~\SI{200}{\micro\meter}\) (crystal thickness), and a voltage ramp of \(dU/dt = \SI{0.25}{V/s}\), we obtain a charging current \(I_{charge} = \SI{0.37}{pA}\), which is larger than the above mentioned offset current. However, the observed discrepancy is reasonable, since practically the voltage sweep is, as said above, not realized in a continuous manner and the charging current shows a certain decay between incrementing the voltage and the respective current recording.

\section{Second Harmonic Generation Microscopy (SHGM)}
\label{sec:app:shg}

\begin{figure}[h!]
    \centering
    \includegraphics[width=0.6\textwidth]{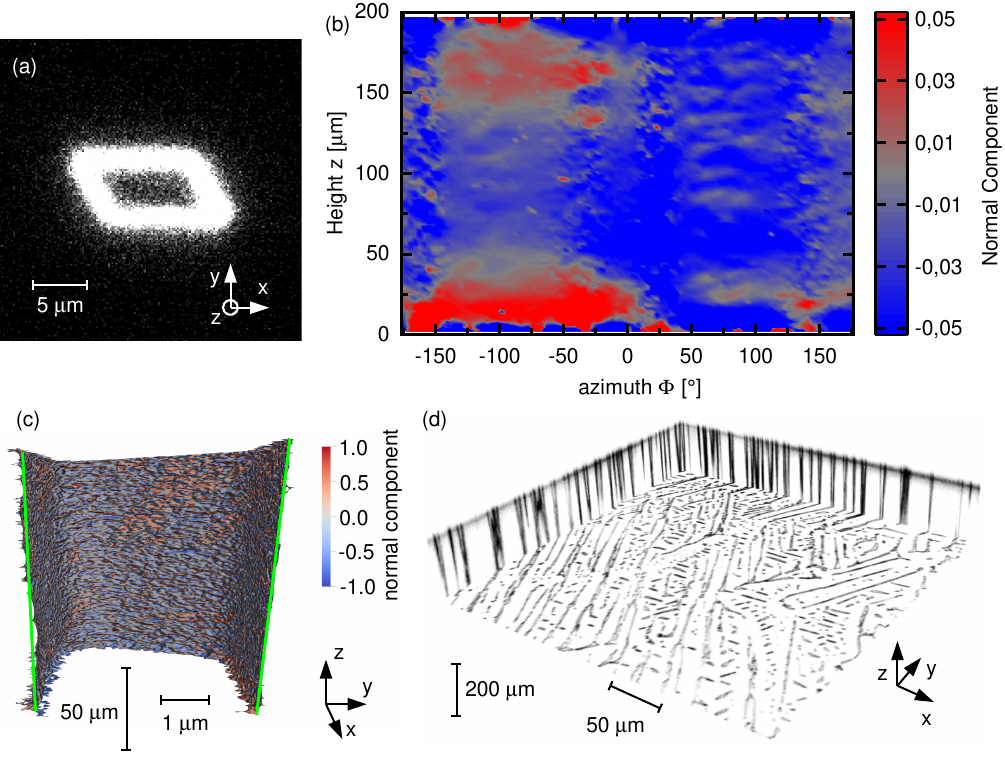}
    \caption{Second harmonic generation microscopy images resolving the 3-dimensional structure of the domain walls in samples \textit{DW-02}~(a--c) and \textit{DW-04}~(d).\\ \textbf{(a)} A cross section in the xy-plane in the middle of the total sample depth shows the parallelogram-like shape of this domain wall. Mostly likely two sides collapsed to zero length and are absorbed in the upper left and lower right corner. \textbf{(b)} By extracting the points of highest intensity, the (close to) right rhombic prism shaped domain wall can be reconstructed. A cut through the yz-plane shows the inclination of the wall with respect the z axis. \textbf{(c)} Orientation of the normal component of the domain wall indicates the inclination angle with respect the z axis. The surface is unfolded in cylindrical coordinates.\\
    \textbf{(d)} Large arrays of spike and through domains are observed in \textit{DW-04} after a kind of irreversible domain "explosion" during the application of the high-voltage ramp for DW conductivity enhancement, as described earlier by Kirbus \emph{et al.}~[14].}
    \label{fig:cshg}
\end{figure}

In order to visualize the 3-dimensional internal structure of the DWs after the conductivity enhancement procedure, we (cf.~SI-\cref{fig:cshg}) performed confocal second harmonic generation microscopy (SHGM) on samples \emph{DW-02} \emph{DW-04} exemplarily. The commercial SHGM setup consisted of two parts: a laser scanning microscope \emph{Zeiss LSM 980} working in tandem with a tunable Ti:Saphire laser \emph{Mai Tai BB, Spectra Physics}. For our investigations, a 100-fs pulse at 900~nm was applied for scanning the sample by a galvanometric xyz-scanner. At the domain wall the laser beam is converted into a specific SHG-signal, which is recorded in back-reflection. The resulting image was captured at a frame rate of 0.1~Hz leading to 1024*1024 pixel images, which were further processed by the software packages \emph{ParaView} and \textit{python3}. For further details, see our previous work~[14].

\newpage
\section{Comparison and testing of different models for activated hopping transport}
\label{sec:app:hopping_models}

In general, so-called \emph{thermally-activated conduction} comprises a number of mechanisms, which have been described in solids with localized electronic states~[35]. The recording of temperature dependent electrical-conduction characteristics allows us to differentiate between the effective transport mechanisms. Thereby, the functional dependence of the conductivity \(\sigma(T)\) is given as:

\begin{equation}
	\sigma(T) =  \tilde{\sigma}_0 T^{- \alpha} \exp \left( - \left[ \frac{T_0}{T} \right]^\beta \right),
	\label{equ:theory:general_conductivity_temperature}
\end{equation}
with \(\tilde{\sigma}_0\) being a constant prefactor determined by the sample geometry, \(T_0\) a characteristic temperature, which can be written via \(T_0 = E_a / k_B\) as an energy \(E_a\), also referred to as (hopping) activation-energy. The dimensionless coefficients \(\alpha\) and \(\beta\) depend on the underlying nanoscopic processes and are therefore the target of any curve fitting procedure. As summarized and explained in much more detail in ref.~[33], three major groups of conduction can be distinguished via the coefficient \(\beta\):

\begin{itemize}
	\item \(\beta = 1\): Thermally-activated hopping~[35,36]. 
	\item \(\beta = \frac{1}{4}\): Mott variable-range hopping~[37]
	\item \(\beta = \frac{1}{2}\): Efros–Shklovskii variable-range hopping~[38]
\end{itemize}

Based on these fundamental mechanisms, various combinations of \(\alpha\) and \(\beta\) can occur for real transport phenomena:

\begin{itemize}
	\item \(\alpha\) = 0, \(\beta\) = 1: Simple thermally-activated hopping
	\item \(\alpha\) = 0, \(\beta = \frac{1}{4}\): Mott variable-range hopping
	\item \(\alpha\) = 0, \(\beta = \frac{1}{2}\): Efros–Shklovskii variable-range hopping
	\item \(\alpha = 1\), \(\beta = 1\): Adiabatic polaron hopping
	\item \(\alpha = \frac{3}{2}\), \(\beta = 1\): Non-adiabatic polaron hopping
	\item \(\alpha = 0.35\), \(\beta = \frac{1}{4}\): Variable-range hopping with constant distance~[39]
	\item \(\alpha = \frac{9}{2}\), \(\beta = \frac{1}{2} \): Efros-Shklovskii hopping with polaronic pseudo-bandgap~[40]
\end{itemize}

Technically, the parameter \(\beta\) dominates the temperature dependence, because it enters exponentially in \cref{equ:theory:general_conductivity_temperature}, whereas \(\alpha\) enters via a power law only. Consequently, \(\alpha\) is hard to be determined and usually complementary experimental techniques, e.g., optical methods like photoluminescence analysis, have to be employed to pinpoint a single transport mechanism.

Returning now to the specific case of \ce{LiNbO3} domain walls -- for more details refer to [34] as well -- single \(I(T)\) curves at constant voltage were analyzed with respect to their compatibility to \cref{equ:theory:general_conductivity_temperature}. Since practically \(\alpha\) and \(\beta\) are highly correlated, it was not possible to fit them simultaneously. As a workaround, one parameter was fixed to a common value, while the other parameter was fitted.

\begin{table}[htb]
	\begin{tabular}{|l|l|r|r|r|r|r|r|} \hline
		\thead{parameter} & \thead{unit} & \thead{\(\alpha\) = 0} & \thead{\(\alpha\) = 1} & \thead{\(\alpha\) = 3/2} & \thead{\(\beta\) = 1} & \thead{\(\beta\) = 1/2} & \thead{\(\beta\) = 1/4} \\ \hline
		\(\alpha\) & - & 0 & 1 & 1.5 & 2.17 & 18.29 & 50.54 \\
		\(\Delta \alpha\) & - & - & - & - & 0.17 & 0.12 & 0.16 \\
		\(\beta\) & - & 1.147 & 1.072 & 1.038 & 1 & 0.5 & 0.25 \\
		\(\Delta \beta\) & - & 0.015 & 0.013 & 0,012 & - & - & - \\
		\(T_0\) & K & 808 & 1020 & 1149 & 1332 & \num{151.3e3} & \num{305.8e8} \\
		\(\Delta T_0\) & K & 25 & 32 & 36 & 14 & \num{1.1e3} & \num{3.2e8} \\
		\(k_B T_0\) & eV & 0.0696 & 0.08786 & 0.099 & 0.1148 & 13.03 & \num{2.6e6} \\
		\(\tilde{I}_0\)  & \(\text{A} \cdot \text{K}^\alpha\) &
			0.00233 & 2.73 & 98 & \num{11.6e3} & \num{1.00e62} & \num{8.12e219} \\
		\(\Delta \tilde{I}_0\)  & \(\text{A} \cdot \text{K}^\alpha\) &
			0.00016 & 0.20 & 7 & \num{5.5e3} & \num{3.71e61} & \num{5.44e219} \\
		\(R^2\) & - & 0.999 & 0.999 & 0.999 & 0.999 & 0.999 & 0.999 \\ \hline
	\end{tabular}
	\caption{Parameters from \(I(T)\)-curve fitting for sample~\textit{DW-01} with \(U = \SI{10}{\volt}\) according to \cref{equ:theory:general_conductivity_temperature}. The fixed values of \(\alpha\) and \(\beta\) represent different models for the intrinsic electrical transport mechanism of a lithium niobate domain wall as listed above.}
	\label{tbl:results:temperatur_conductivity_models}
\end{table}

The fit parameters are summarized in \cref{tbl:results:temperatur_conductivity_models}. All fit curves agree with the measured data according to their $R^2$~coefficient nearly perfectly, so a closer analysis is required. It further proves the high correlation of \(\alpha\) and \(\beta\) that prevents the distinction of their influences.

Most obviously, the cases of fixed \(\beta\) with \(\beta = \frac{1}{2}\) and \(\beta = \frac{1}{4}\) can be discarded, since the values of \(\alpha\) and \(T_0\) are orders of magnitudes away from all predictions of any model found in the literature. Moreover, all approaches with fixed \(\alpha\) converged with \(\beta\) being close to one, despite the estimated uncertainties of \(\beta\) are probably underrated. Thus, thermally-activated hopping with \(\beta = 1\) is the most reasonable conduction mechanism behind the observed domain wall currents, which indicates an equal energy of the (polaron) hopping sites, while there is no statement about their real space distribution possible. No indications for variable-range hopping as introduced in~[38] or [39] could be found.

In a next step, the determination of \(\alpha\), which allows for a detailed distinction between the different mechanisms, is tested. However, this attempt failed due to the fact that \(\beta\) influences the conductivity \(\sigma\) exponentially, while \(\alpha\) interferes quite weakly via a power law only. Consequently, all curve fittings with fixed \(\alpha\) converge and result in \(\beta \approx 1\). For the possible case of small free polarons with \(\alpha = \frac{3}{2}\), \(\beta\) converges indeed with the lowest difference to \num{1.0} among all fixed-\(\alpha\) fits, but it is not statistically evident.

In conclusion, we proceed using the case of simple thermally-activated hopping transport for all further data processing described in the main text.

\section{Current-voltage (IU) diagrams in logarithmic representation}
\begin{figure}[h!]
    \centering
    \includegraphics[width=0.55\textwidth]{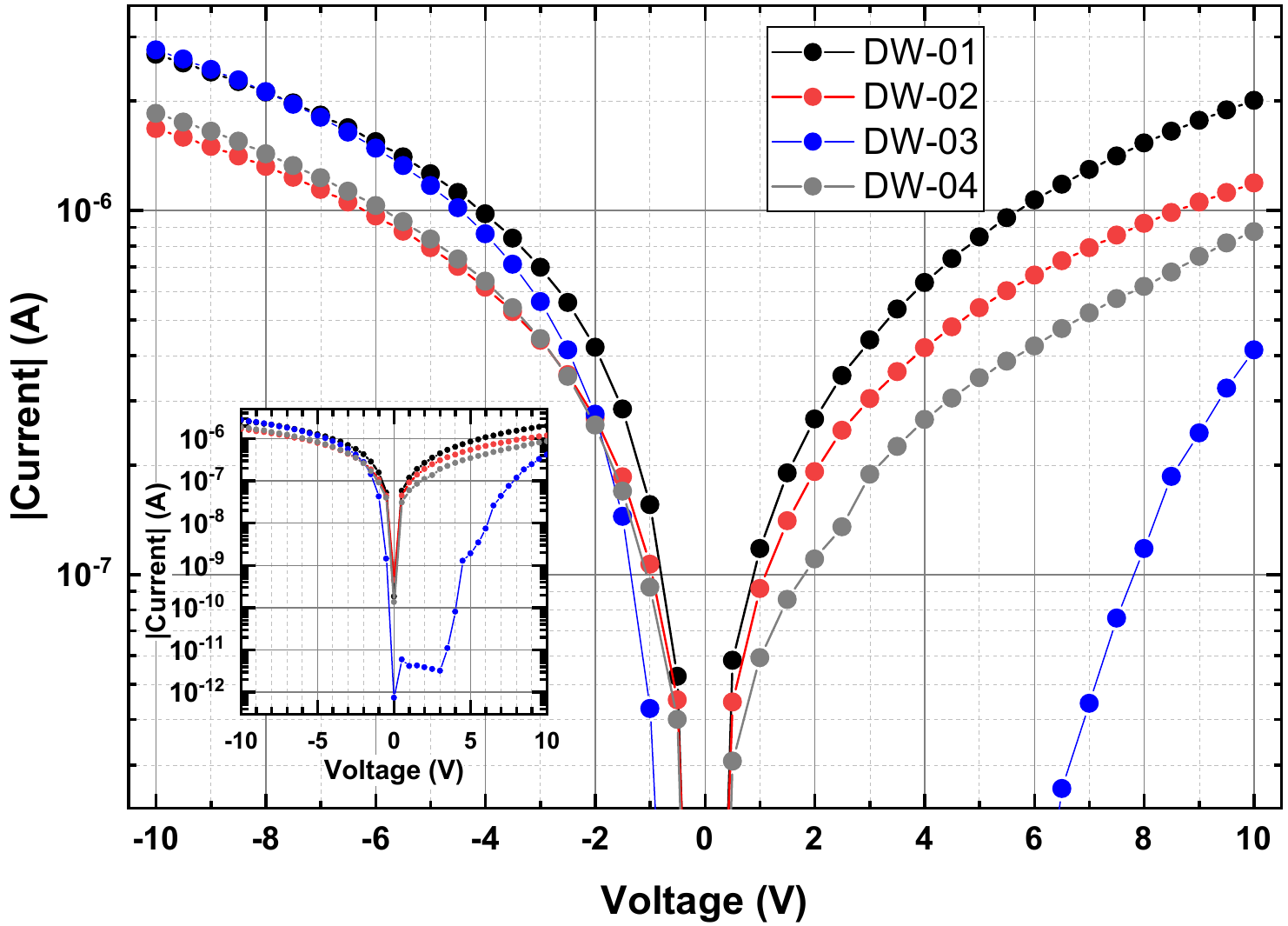}
    \caption{Logarithmic representation of the IU characteristics of fig.~4 of the main text: full curves are shown in the inset, a zoomed version in the main picture.}
    \label{fig:RT_IU_log}
\end{figure}

SI-\cref{fig:RT_IU_log} displays the same data sets as fig.~4 of the main text, but with logarithmically plotted (absolute values of the) current, which allows a clearer view towards the asymmetry of the current with the polarity.

\section{Temperature dependence of the diode ideality factors}
\label{sec:app:n_T}

\begin{figure}[h!]
    \centering
    \includegraphics[width=0.5\textwidth]{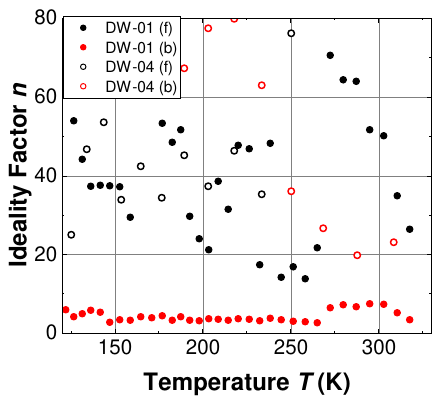}
    \caption{Temperature dependence of the diode ideality factors of samples DW-01 and DW-04.}
    \label{fig:n_T}
\end{figure}

\Cref{fig:n_T} displays the temperature dependence of the ideality factor \(n\) for \emph{DW-01} and \emph{DW-04} for both voltage directions and complements fig.~5 of the main paper. At a first glance, the $n$~values exhibit strong fluctuations. Only for \textit{DW-01} a rather constant value is observed in backward direction, which would be the expected result, since according to the thermionic emission model described in~[31], $n$ depends on the effective doping concentration that corresponds to the density of hopping sites, being independent of temperature. In summary, it can be stated that (i) $n$ is always much larger than one and (ii) is the most unstable parameter. 

\section{R2D2 fit parameters and their uncertainty}
\label{sec:app:fit_parameter}


The following section comprises three tables, in which the fit parameters including their statistical errors for all samples that were considered in the main text are summarized.

First, \cref{tab:dc-parameter-uncertainties} shows the fit parameters and -- as an addition to table~I of the main text -- their uncertainties for the four DWs in bulk LNO samples measured at room temperature, i.e., \emph{DW-01...DW-04}. 

\begin{table}[ht]
    \setlength{\tabcolsep}{0.1em}
    \begin{tabular}{| l
			| >{\collectcell}r<{\endcollectcell}
			@{\({}\pm{}\)}
			>{\collectcell}l<{\endcollectcell} |
			>{\collectcell}r<{\endcollectcell}
			@{\({}\pm{}\)}
			>{\collectcell}l<{\endcollectcell} |
			>{\collectcell}r<{\endcollectcell}
			@{\({}\pm{}\)}
			>{\collectcell}l<{\endcollectcell} | r |} \cline{1-7}

        \thead{sample} & \multicolumn{2}{c|}{\thead{\(R_f\) [\si{\mega\ohm}]}} & \multicolumn{2}{c|}{\thead{\(I_{s, f}\) [pA]}}
		 & \multicolumn{2}{c|}{\thead{\(n_f\) [-]}} & \multicolumn{1}{r}{} \\ \cline{1-7}
		\textit{DW-01} & 3.68 & 0.12 & (1.23 & 0.34) \(\cdot \num{e5}\) & 35.7 & 6.3 & \multicolumn{1}{r}{} \\
		\textit{DW-02} & 7.156 & 0.034 & (9.83 & 0.14) \(\cdot \num{e4}\) & 23.6 & 0.2 & \multicolumn{1}{r}{} \\ 
		\textit{DW-03} & 2.838 & 0.005 & 12.8 & 0.7 & 33.64 & 0.16 & \multicolumn{1}{r}{} \\
		\textit{DW-04} & 2.59 & 0.18 & (3.41 & 0.04) \(\cdot \num{e5}\) & 240 & 5 \\
		\cline{1-7}

		\multicolumn{6}{l}{} \\
		\hline
		\thead{sample} & \multicolumn{2}{c|}{\thead{\(R_b\) [\si{\mega\ohm}]}} & \multicolumn{2}{c|}{\thead{\(I_{s, b}\) [pA]}}
		& \multicolumn{2}{c|}{\thead{\(n_b\) [-]}} & \(R^2\) \\ \hline
		\textit{DW-01} & 3.437 & 0.008 & 430 & 290 & 5.3 & 0.9 & 0.990 \\
		\textit{DW-02} & 5.511 & 0.010 & 100 & 90 & 5.3 & 0.5 & 0.980 \\
		\textit{DW-03} & 3.06987 & 0.00038 & 211.4 & 2.0 & 6.407 & 0.008 & 0.951 \\
		\textit{DW-04} & 4.482 & 0.024 & ( 4.91 & 0.30 ) \(\cdot \num{e3}\) &
		    19.8 & 0.4 & 0.957 \\
		\hline
	\end{tabular}
    \caption{Full set of equivalent circuit (R2D2) parameters for the IU characteristics of the four single-crystal-based DWs shown in fig.~4 and table~I \emph{including} their corresponding uncertainties $R^2$. Note that the latter ones all lie close to 1. }
    \label{tab:dc-parameter-uncertainties}
\end{table}

Second, \cref{tab:parameter_TFLN} shows the R2D2 fit parameters for the thin film LNO, poled with an atomic force microscopy tip, which is discussed in sec.~III~C of the main paper. 

\begin{table}[h!]
    \center
    \setlength{\tabcolsep}{0.1em}
    \begin{tabular}{
			| >{\collectcell}r<{\endcollectcell}
			@{\({}\pm{}\)}
			>{\collectcell}l<{\endcollectcell} |
			>{\collectcell}r<{\endcollectcell}
			@{\({}\pm{}\)}
			>{\collectcell}l<{\endcollectcell} |
			>{\collectcell}r<{\endcollectcell}
			@{\({}\pm{}\)}
			>{\collectcell}l<{\endcollectcell} |
			>{\collectcell}r<{\endcollectcell}
			@{\({}\pm{}\)}
			>{\collectcell}l<{\endcollectcell} |
			>{\collectcell}r<{\endcollectcell}
			@{\({}\pm{}\)}
			>{\collectcell}l<{\endcollectcell} |
			>{\collectcell}r<{\endcollectcell}
			@{\({}\pm{}\)}
			>{\collectcell}l<{\endcollectcell} |
			r |} 

		\hline
		\multicolumn{2}{|c|}{\thead{\(R_f\) [\si{\mega\ohm}]}} &
		\multicolumn{2}{c|}{\thead{\(I_{s, f}\) [pA]}} &
		\multicolumn{2}{c|}{\thead{\(n_f\) [-]}} &
		\multicolumn{2}{c|}{\thead{\(R_b\) [\si{\mega\ohm}]}} &
		\multicolumn{2}{c|}{\thead{\(I_{s, b}\) [pA]}} &
		\multicolumn{2}{c|}{\thead{\(n_b\) [-]}} & \(R^2\) \\ \hline
		17.5 & 0.1 & 16 & 1 & 23.2 & 0.2 & 94 & 9 & 507 & 74 & 76 & 5 & 0.99995 \\ \hline
	\end{tabular}
	\caption{Equivalent circuit parameters as extracted by applying the R2D2 model to the IU curve of a z-cut thin film LNO domain wall shown in fig.~7.}
	\label{tab:parameter_TFLN}
\end{table}

Third, as also discussed in sec.~III~C, we applied the model to the IU-curves published Qian~\emph{et~al.}~[19] in order to prove the validity of the R2D2 model on an even larger range of data. There, we extracted the R2D2 fit parameters and their uncertainties as shown in table~S4. 

\begin{table}[h!]
    \center
    \setlength{\tabcolsep}{0.1em}
    \begin{tabular}{| l
			| >{\collectcell}r<{\endcollectcell}
			@{\({}\pm{}\)}
			>{\collectcell}l<{\endcollectcell} |
			>{\collectcell}r<{\endcollectcell}
			@{\({}\pm{}\)}
			>{\collectcell}l<{\endcollectcell} |
			>{\collectcell}r<{\endcollectcell}
			@{\({}\pm{}\)}
			>{\collectcell}l<{\endcollectcell} |
			>{\collectcell}r<{\endcollectcell}
			@{\({}\pm{}\)}
			>{\collectcell}l<{\endcollectcell} |
			>{\collectcell}r<{\endcollectcell}
			@{\({}\pm{}\)}
			>{\collectcell}l<{\endcollectcell} |
			>{\collectcell}r<{\endcollectcell}
			@{\({}\pm{}\)}
			>{\collectcell}l<{\endcollectcell} |
			r |} 

		\hline
		\thead{sample} &
		\multicolumn{2}{c|}{\thead{\(R_f\) [\si{\giga\ohm}]}} &
		\multicolumn{2}{c|}{\thead{\(I_{s, f}\) [pA]}} &
		\multicolumn{2}{c|}{\thead{\(n_f\) [-]}} &
		\multicolumn{2}{c|}{\thead{\(R_b\) [\si{\giga\ohm}]}} &
		\multicolumn{2}{c|}{\thead{\(I_{s, b}\) [pA]}} &
		\multicolumn{2}{c|}{\thead{\(n_b\) [-]}} & \(R^2\) \\ \hline
		\textit{head-08} & 3.8 & 0.5 & 81 & 8 & 81.7 & 3.0 & 5.2 & 0.5 & 0.83 & 0.28 & 17.2 & 1.2 & 0.960 \\
		\textit{head-10} & 3.67 & 0.11 & 34.8 & 0.5 & 33.16 & 0.10 & 3.67 & 0.31 & 3.6 & 0.9 & 14.0 & 0.9 & 0.961 \\
		\textit{tail-08} & 33.1 & 2.3 & 2.93 & 0.10 & 202.8 & 3.1 & 48.4 & 1.8 & 0.129 & 0.006 & 107.98 & 0.24 & 0.967 \\ 
		\textit{tail-10} & 40 & 9 & 4.76 & 2.6 & 192 & 43 & 5 & 12 & 11.3 & 0.8 & 335 & 32 & 0.933 \\ \hline
	\end{tabular}
	\label{tab:parameter_qian_2022}
	\caption{Equivalent circuit parameters as extracted by applying the R2D2 model to the IU curves observed by Qian~\textit{et al.} [19] on x-cut thin film LNO domain walls.}
\end{table}


\section{Derivation of the voltage dependence of the activation energy}
\label{sec:app:ea_vs_v}

In addition to the R2D2 model, we directly determined activation energies from the temperature-dependent IU measurements. This approach is not as straightforward as the R2D2 model, because the obtained energies at low voltages are effectively a convolution of the resistor and diode parameters. At high voltages, the potential mainly drops over the resistor, leading to similar values for the activation energy as observed for the pure resistive contribution within the R2D2 model. At low voltages, the potential drops over the diode that dominates then also the activation energy observed. A numerical model for this effective activation energy can be derived by calculating the partial derivative of the logarithm of the current with respect to the inverse temperature:

\begin{equation}
	\begin{aligned}
		E_a & := - k_B \frac{\partial \ln I / I_0}{\partial 1/T}
            = - k_B \frac{\partial \ln I}{\partial 1/T}
            + k_B \frac{\partial \ln I_0}{\partial 1/T}
            = - k_B \frac{\partial \ln I}{\partial 1/T} \\
		& {=}
 		 - k_B \frac{\partial}{\partial 1/T} \ln \left\{
			A^\star T^2 \exp \left( \frac{- q \Phi_{eff}}{k_B T} \right)
		\left[ \exp \left( \frac{q U}{n k_B T} \right) -1 \right] \right\} \\
		& = - k_B \frac{\partial}{\partial 1/T} \ln(A^\star T^2) 
			- k_B \frac{\partial}{\partial 1/T} \left( 
			\frac{-q \Phi_{eff}}{k_B T} \right)\\
		& \qquad - k_B \frac{\partial}{\partial 1/T} \ln \left[ 
			 \exp \left( \frac{q U}{n k_B T} \right) - 1 \right] \\
		& = \underbrace{2 k_B T + q \Phi_{eff}}_{:= E_0}
			- \underbrace{k_B T}_{:= A}
				\underbrace{\frac{q}{n k_B T}}_{:= 1/U_c} U
				\frac{1}{1 - \exp (- \frac{q U}{n k_B T})} \\
		& = E_0 - A \cdot \frac{U/U_c}{1 - \exp(-U/U_c)}. \\
	\end{aligned}	
	\label{equ:activation_energy_complete}
\end{equation}

A shown in fig.~6, the parameters $E_0$, $A$, and $U_c$ can be extracted such that the model describes the experimental data well, while the physical interpretation of these parameters is not straightforward.

\section{Experimental details on domain walls in thin film lithium niobate: preparation~and~recording~of~IU~characteristics}
\label{sec:app:thin_film_preparation}

The thin-film samples used were \SI{500}{nm} thick, single crystal, z-cut \ce{LiNbO3}, with a \SI{150}{nm} Au-Cr bottom electrode, obtained from \textit{NanoLN}. Poling was achieved by application of super-coercive voltage pulses ($\approx \SI{50}{V}$) to an AFM tip, during the course of a contact mode scan. A specialized tip holder with high voltage capabilities was used in conjunction with an \textit{MFP-3D Infinity} AFM system. Arrays of approximately conical, \SI{180}{\degree} domains with slightly inclined (charged) and conducting domain walls were achieved, similar to results presented in previous work by McCluskey \emph{et al.}~[52] -- the citation shows examples of the typical microstructures we get by AFM poling.These were confirmed by piezoresponse force microscopy and conducting-AFM with sub-coercive voltages.

Planar, \SI{100}{nm} thick silver electrodes, measuring approximately $100 \times \SI{100}{\mu m^2}$, were thermally evaporated onto the multidomain area. Current-voltage measurements were performed on the resultant capacitor structures by the application of voltage pulses to a tungsten microprobe in contact with the silver top electrode, using a \textit{Keysight~B2910BL} Source/Measure unit. The voltage was swept from \SI{0}{V} to \SI{+10}{V}, to \SI{-10}{V} and back to \SI{0}{V}, in steps of \SI{0.1}{V}. The pulse length was \SI{0.1}{s}. A small droplet of liquid indium-gallium-tin eutectic was placed between the probe and the electrode to maintain good electrical contact. The circuit ground was connected to the bottom electrode of the \ce{LiNbO3} by conducting silver paste.

\end{document}